\documentclass[11pt]{article}
\usepackage{amssymb,latexsym,amsmath,amsthm,graphicx,subfigure}

\makeatletter \@addtoreset{figure}{section}
\def\thefigure{\thesection.\@arabic\c@figure}
\def\fps@figure{h, t}
\@addtoreset{table}{bsection}
\def\thetable{\thesection.\@arabic\c@table}
\def\fps@table{h, t}
\@addtoreset{equation}{section}

\newtheorem{theorem}{Theorem}[section]
\newtheorem{proposition}[theorem]{Proposition}

\newfont{\tenbi}{cmbxti10}

\newcommand{\bs}{\boldsymbol}
\newcommand{\Tr}{{\rm Tr}}

\begin{document}
\title{Chaplygin ball over a fixed sphere: explicit integration
\footnote{AMS Subject Classification 37J60, 37J35, 70H45}}

\author{A. V.\,Borisov \\ 
Institute of Computer Science, Udmurt State University, \\
ul. Universitetskaya 1, Izhevsk, 426034 Russia \\
e-mail: {\tt borisov@rcd.ru} \\ and \\
Yu. N. Fedorov \\
 Department de Matem\`atica I, \\ 
Universitat Politecnica de Catalunya, \\
Barcelona, E-08028 Spain \\
e-mail: {\tt Yuri.Fedorov@upc.edu} \\ and \\
I. S.\,Mamaev \\ 
Institute of Computer Science, Udmurt State University, \\
ul. Universitetskaya 1, Izhevsk, 426034 Russia \\
e-mail: {\tt mamaev@rcd.ru} }
\date{}

\maketitle

\abstract{We consider a nonholonomic system describing a rolling of a
dynamically non-symmetric sphere over a fixed sphere without slipping. The system generalizes the
classical nonholonomic Chaplygin sphere problem and it is shown to be integrable for one special ratio
of radii of the spheres. After a time reparameterization the system becomes a Hamiltonian one and
admits a separation of variables and reduction to Abel--Jacobi quadratures.
The separating variables that we found appear to be a non-trivial generalization of ellipsoidal
(spheroconic) coordinates on the Poisson sphere, which can be useful in other integrable problems.

Using the quadratures we also perform an explicit integration of the problem in theta-functions of the new time. }

\section{Introduction}
One of the best known integrable systems of the classical
nonholonomic mechanics is the Chaplygin problem on a
non-homogeneous sphere rolling over a horizontal plane without
slipping. In \cite{9} S.\,A. Chaplygin obtained the equations of
motion, proved their integrability and performed their reduction
to quadratures by using spheroconical coordinates on the Poisson sphere as separating variables.
He also actually solved the reconstruction problem by
describing the motion of the sphere on the plane.

Various aspects of this celebrated system were studied in
\cite{2,5,9,11}, and its explicit integration in terms of
theta-functions was presented in \cite{Ch_sol}.

Several nontrivial integrable generalizations of this problem were
indicated by V. Kozlov \cite{Kozlov1985} (the motion of the sphere
in a quadratic potential field), A. Markeev \cite{Markeev1986}
(the sphere carries a rotator), in \cite{Ves} (an extra
nonholonomic constraint is added) and in \cite{Fed_support} (the
sphere touches an arbitrary number of symmetric spheres with fixed
centers).

Next, amongst others, the papers \cite{1,2, Yaroshuk} considered rolling of the
Chaplygin (i.e., dynamically non-symmetric)
sphere over a fixed sphere, so called {\it sphere-sphere problem}. They studied the
equations of motion in the frame attached to the body. More generic (although more tedious) form
of the equations also appeared in the works of Woronetz \cite{29, 292}, who, nevertheless, solved a series
of interesting problems describing rolling of bodies of revolution or flat bodies over a sphere.

A rolling of a generic convex body over a sphere was also discussed in the recent survey \cite{28}.

In \cite{1,2, Yaroshuk} it was observed that the equations of motion of the Chaplygin sphere-sphere problem
admit an invariant measure, but, as numerical computations show,
in the general case they are not integrable (there is no analog of
the linear momentum integral). However, as was also found in \cite{1},
for one special ratio of radii of the two spheres an analog
of such an integral does exist, and the system in integrable by
the Jacobi last multiplier theorem.

Until recently, no one of the above generalizations was integrated
in quadratures (except a case of particular initial conditions of
the Kolzov generalization, considered in \cite{Fed_0}).

In this connection it should be noted that a Lax pair with a spectral parameter for
the Chaplygin sphere problem or for its generalizations is still
unknown and, probably does not exists. Hence, one cannot use the
powerful method of Baker--Akhieser functions to find
theta-function solution of the problem.

\paragraph{Contents of the paper.}
Our main purpose is to find appropriate separating variables, which allow to reduce the integrable case of the
Chaplygin sphere-sphere problem to quadratures, as well as to give explicit theta-function solution.

It appears that, in contrast to
the classical Chaplygin sphere problem, the usual spheroconical coordinates on the Poisson
sphere do not provide separating variables and that such variables should be introduced in a more
complicated way (see formulas (\ref{33})).

Using the quadratures, we also give a brief analysis of possible bifurcations and periodic solutions.
These results are presented in Sections 3-4.

In Section 5 we briefly describe another type of periodic solutions.

Section 6 provides a derivation of explicit theta-function solutions of the problem in a self-contained form 
(Theorem \ref{main}).

Finally, in Appendix we show how the separating variables we used can be obtained in a systematic way,
by reducing a restriction of our system to an integrable Hamiltonian system with 2 degrees of freedom and
applying a classical result of Eisenhart on transformation of the Hamiltonian to a St\"ackel form.

\section{Equations of motion and first integrals}
Consider rolling of the Chaplygin sphere inside/over a fixed sphere without slipping (Figure 1.)

Let $\boldsymbol\omega$, $m$, ${\bf I}={\rm diag}(I_1,\,I_2,\,I_3)$, and $b$ denote respectively the angular
velocity vector of the Chaplygin sphere, the mass of the sphere, its inertia tensor, and the radius.

By ${\boldsymbol n}$ we denote the unit normal vector to the fixed sphere $S^2$ at the contact point $P$.
The angular momentum ${\boldsymbol M}$  of the moving sphere with respect to $P$ can be written as
\begin{equation} \label{mom}
{\boldsymbol M}={\bf I}{\boldsymbol\omega}
 + d\, {\boldsymbol n}\times ({\boldsymbol n}\times{\boldsymbol\omega}),\qquad d=mb^2,
\end{equation}
where $\times$ denots the vector product in ${\mathbb R}^3$.

The phase space of the dynamical system is the tangent bundle $T(SO(3)\times S^2)$.
By using the no slip nonholonomic constraint (which
corresponds to zero velocity in the point of contact),
one can obtain the reduced equations of motion in the frame attached to the sphere in the following
closed form (see, e.g., \cite{1}):
\begin{equation}
\label{1}
\dot{\boldsymbol M} ={\boldsymbol M}\times\boldsymbol\omega,\quad
\dot {\boldsymbol n} =k\, {\boldsymbol n}\times\boldsymbol\omega,\quad k=\frac{a}{a+b},
\end{equation}
$a$ being the radius of the fixed sphere.

Note that the ratio $k$ can take any positive or negative value depending on the relative position of the rolling and
fixed spheres, as illustrated in Fig.~1.
\begin{figure}
\centering\includegraphics{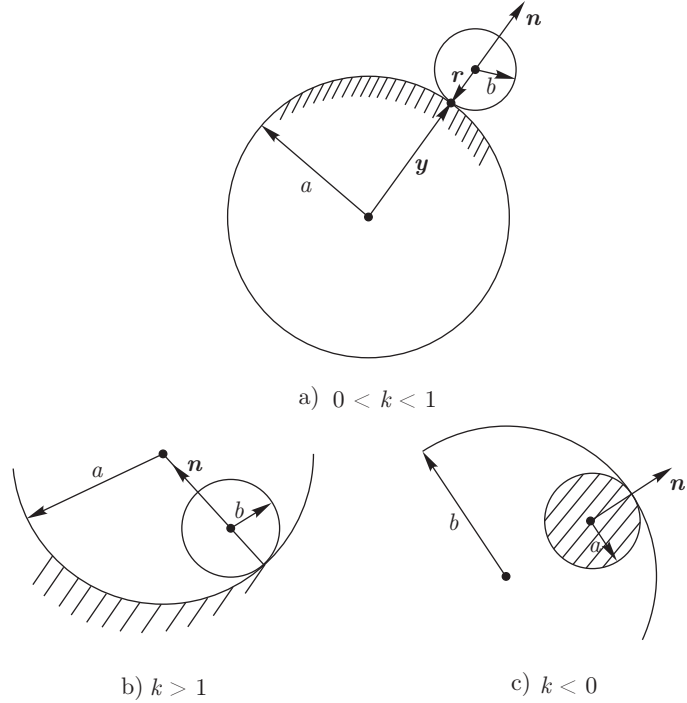}
\caption{Rolling of the Chaplygin sphere inside/over the fixed sphere (dashed).}
\end{figure}

For arbitrary $k$ the equations \eqref{1} possess three independent integrals
\begin{equation}
\label{2}
F_0=\langle{\boldsymbol n},\,{\boldsymbol n}\rangle=1,\quad
H= \langle{\boldsymbol M},\,{\boldsymbol\omega}\rangle,\quad
F_1=\langle{\boldsymbol M}, \,{\boldsymbol M}\rangle ,
\end{equation}
which, in view of \eqref{mom}, can be written as
\begin{equation} \label{H,F}
\begin{aligned}
F_0 & =\langle{\boldsymbol n},\,{\boldsymbol n}\rangle=1,\quad
H= \langle \bs\omega, {\bf J}\bs\omega \rangle - d^2 \langle \bs n , \bs\omega \rangle^2 ,\\
F_1 & = \langle {\bf J} \bs\omega, {\bf J}\bs\omega \rangle -2\langle {\bf J} \bs\omega, \bs n \rangle\, \langle \bs n , \bs\omega \rangle
 + d^2 \langle \bs n, \bs\omega \rangle^2,
\end{aligned}
\end{equation}
where ${\bf J}={\bf I}+d{\bf E}$, $\bf E$ being the identity matrix.

As shown in \cite{Yaroshuk}, the equations \eqref{1} expressed in terms of $omega, n$ also have
the invariant measure $\rho\,d{\boldsymbol \omega}\,d{\boldsymbol n}$ with the density
\begin{equation}
\label{density}
\rho = \sqrt{\langle{\boldsymbol n},\,{\boldsymbol n}\rangle
-d\langle{\boldsymbol n},\,{\bf J}^{-1}{\boldsymbol n}\rangle}.
\end{equation}
Thus, according to the Jacobi theorem, for a complete integrability of this system one extra integral is needed.

\paragraph{The Chaplygin sphere on the plane.}
Clearly, the case $k=1$ corresponds to $a\to\infty$, that is, the fixed sphere transforms to a horizontal plane
with the unit normal vector $n$,
and we arrive at the classical integrable Chaplygin problem, when the linear (in $M$) momentum integral is preserved:
\begin{equation}
\label{lin}
\langle {\boldsymbol M}, {\boldsymbol n} \rangle \equiv \langle {\bf I}{\bs\omega}, {\boldsymbol n} \rangle.
\end{equation}

\paragraph{Second integrable case.}
According to~\cite{1},  the system ~\eqref{1} is also integrable in the case $k=-1$, which describes
rolling of a non-homogeneous ball with a spherical cavity over a fixed sphere
and the quotient of the radii of the spheres equals $\frac{b}{a}=\frac{1}{2}$ (see Fig. 1 c).

In this case, instead of \eqref{lin}, there is the following linear
integral\footnote{We could not interpret this integral as a momentum conservation law.}
\begin{equation}
\label{22}
F_2= \langle {\bf A} {\boldsymbol M}, \, {\boldsymbol n} \rangle\, ,
\end{equation}
where
$$
{\bf A}={\rm diag}( J_2+J_3-J_1,\,J_3+J_1-J_2,\,J_1+J_2-J_3)\, .
$$

Note that, as was shown in \cite{16},
the modification of this system obtained by imposing the extra ``no twist'' constraint
$\langle \bs\omega,{\boldsymbol n}\rangle=0$ (sometimes called as {\it the rubber Chaplygin ball})
is also integrable for the ratio $k=-1$.

In the next sections we present explicit integration of this case under the condition $F_2=0$.
Our procedure is similar to that of the problem of the Chaplygin sphere rolling on a horizontal plane
in the case of zero value of the area integral \eqref{lin}, (see~\cite{9,5,8}),
however, analytically, it is more complicated.

Integration of the system \eqref{1} with $k=-1$ in the general case $F_2\neq 0$ is still an open problem.

\paragraph{A remark on reduction to quadratures in the case $d=0$.}
Note that in the limit case $d=0$ one has ${\bs M}= {\bf I} \bs\omega$ and the equations
\eqref{1} with $k=-1$  take the form
$$
{\bf I}\dot{\bs\omega} = {\bf I} \bs\omega \times{\bs\omega}, \quad
\dot{\boldsymbol n}  = - {\boldsymbol n} \times{\bs\omega}.
$$
As was noticed in \cite{1}, by the substitution
\begin{equation}
\label{subs}
{\bs {\mathcal M}}={\bf AI}{\bs\omega}, \quad {\bs\gamma}= {\boldsymbol n}
\end{equation}
and the sign change $t \to -t$,
the latter system transforms to the Euler--Poisson equations for the classical Euler top,
\begin{gather}
\label{E-P}
\dot{\bs{\mathcal M}} = {\bs{\mathcal M}} \times {\bs\omega}, \quad \dot  {\bs\gamma} = {\bs\gamma}\times{\bs\omega},\\
\omega_i =a_i {\mathcal M}_i, \quad  a_i = \frac {1}{(J_j+J_k-J_i)J_i}, \label{a_i}
\end{gather}
which possesses first integrals
$$
\langle {\bs{\mathcal M}}, {\bs\gamma} \rangle=g,  \quad \langle {\bs{\mathcal M}}, {\bs{\mathcal A}} {\bs{\mathcal M}}\rangle =h , \quad
\langle {\bs{\mathcal M}}, {\bs{\mathcal M}}\rangle =f ,
$$
where $ {\bs{\mathcal A}} =\rm{diag}(a_1,a_2,a_3)$.

As was indicated in several publications (see, e.g., \cite{Acta_bill}),
the Euler--Poisson equations (\ref{E-P}) can be integrated
by separation of variables. Namely, by an appropriate choice of the constant vector $\bs \gamma$ in space
we can always set $g=0$. Then $\langle {\bs{\mathcal A}} \omega, {\bs\gamma}\rangle=0$, and
the equations \eqref{E-P} reduce to a flow on the tangent bundle of the Poisson sphere
$S^2=\{ \langle {\bs\gamma}, {\bs\gamma} \rangle =1 \}$.
In the spheroconical coordinates $\lambda_1, \lambda_2$ on $S^2$ such that
\begin{equation}
\label{lambdas}
\gamma_i^2= \frac{(a_i-\lambda_1)(a_i-\lambda_2 )}{(a_i-a_j)(a_i- a_k)},\qquad i\neq j\neq k\neq i,
\end{equation}
the flow is reduced to the quadratures
\begin{gather}
\label{A-J-0}
\begin{aligned}
\frac {d \lambda_1}{ \sqrt {R(\lambda_1) } } + \frac {d\lambda_2}{ \sqrt {R(\lambda_2) } } & = 0, \\
\frac {\lambda_1\, d\lambda_1}{ \sqrt {R (\lambda_1) } } + \frac
{\lambda_2\, d\lambda_2}{\sqrt { R (\lambda_2)}}&= C\, dt, \quad C=\mbox{const} ,
\end{aligned} \\
R(\lambda)= -(\lambda- a_1)(\lambda- a_2)(\lambda- a_3) (f\lambda- h). \nonumber
\end{gather}
The latter contain one holomorphic and one meromorphic differential on the elliptic curve
${\cal E}= \{\mu^2=R(\lambda)\}$. Thus the the quadratures give rise to
a generalized Abel--Jacobi map and,
following the methods developed in \cite{Cl_Gor},
they can be inverted to express the variables $\bs \gamma, \bs \omega$ in terms of theta-functions of
${\cal E}$ and exponents (see, e.g., \cite{Jac_corps, Acta_bill} for the concrete expressions).
\medskip

Apparently, in the general case $d\ne 0$ the substitution (\ref{subs}) is not useful to integrate the
the equations \eqref{1} in the second integrable case $k=-1$. In particular, it does not
transform these equations to the case of the classical Chaplygin sphere problem ($k=1$).

\section{Reduction to quadratures in the case $F_2=0$}
We now consider the case $d\ne 0$, but assume that the linear integral $F_2$ in \eqref{22} is zero,
which imposes restrictions of the initial conditions. Then, from ~\eqref{1} with $k=-1$ and
$F_2=0$ we get
$$
\dot {\boldsymbol n} = - {\boldsymbol n}\times\boldsymbol\omega, \quad
\langle \bs\omega, {\bf B}\bs n\rangle = 0,
$$
where ${\bf B}=({\bf J}-d{\boldsymbol n}\otimes{\boldsymbol n}){\bf A}$. This allows to
express the angular velocity in terms of $\dot {\boldsymbol n}, {\boldsymbol n}$ in the following homogeneous form
\begin{equation}
\label{3}
{\boldsymbol\omega}=\frac{{\bf B}{\boldsymbol n}\times\dot{\boldsymbol n}}
{ \langle  {\boldsymbol n}, \,{\bf B}{\boldsymbol n}\rangle } .
\end{equation}

In view of the above remark on the reduction to quadratures in the case $d=0$, to perform separation of
variables it seems natural to use the
spheroconical coordinates $\lambda_1,\, \lambda_2$ given by \eqref{lambdas} (with $\gamma_i$ replaced by $n_i$)
in the general case $d\ne 0$ too. However, this choice does not lead to success: after some calculations one can see that
the first integrals $H,F_1$ have mixed terms in the derivatives $\dot\lambda_1,\, \dot\lambda_2$.

It appears that a correct choice is given by
the following {\it quasi-spheroconical} coordinates $z_1,\,z_2$ on the Poisson sphere
$\langle {\boldsymbol n}, {\boldsymbol n} \rangle=1$:
\begin{gather} \label{33}
n_i^2= \frac 1 {G(z_1,z_2)} \frac{\det {\bf I}}{(J_i-d) J_j J_k}\,
\frac{(a_i- z_1)(a_i- z_2 )}{(a_i-a_j)(a_i- a_k)}\, , \qquad (i, j, k)=(1,2,3),
\end{gather}
where
\begin{gather}
G (z_1, z_2) = 1 - d (\Tr {\bf J} -2 d)(z_1+z_2)+ d (4 \det {\bf J}- d \, \Tr( {\bf JA}) ) z_1 z_2 , \label{G}
\end{gather}
and, as in \eqref{a_i}, $a_i= (A_i J_i)^{-1}$.
A systematic derivation of the substitution \eqref{33} is presented in Appendix.

Note that when $d=0$, the factor $G$ becomes 1 and the relation \eqref{33} takes the form of
\eqref{lambdas}, that is, $z_1,z_2$ do become the usual spheroconical coordinates on $S^2$.

We note that similar quasi-spheroconical coordinates were already used in \cite{16, 28}
to integrate the "rubber" Chaplygin sphere-sphere problem.

In the above coordinates $z_1,z_2$ one has
\begin{equation}
\label{subs1}
\rho^2 = \langle ({\bf I} \bs n, {\bf J}^{-1} \bs n\rangle = \frac {\det {\bf I}}{\det {\bf J}}
\frac 1 {G(z_1, z_2)} , \quad
\langle \bs n, {\bf B} \bs n\rangle = \det {\bf I} \det {\bf A}\, \frac {z_1 z_2} {G(z_1, z_2)}
\end{equation}
and
\begin{equation}
\label{dot_n}
\dot n_i = \frac 12 \left ( \frac{\dot z_1}{z_1-a_i}+ \frac{\dot z_2}{z_2-a_i} -
\frac {\dot G (z_1, z_2)} {G(z_1, z_2)} \right) n_i \, .
\end{equation}
Then the expressions \eqref{3} yield
\begin{gather}
\label{om's}
\omega_i = \frac {n_j n_k } 2 \frac{ J_j-J_k }{J_i-d}
\left[ \frac{(1-d A_i z_2)  \dot z_1 }{ (1- a_j^{-1} z_1 ) (1- a_k^{-1} z_1) z_2}+
\frac{(1-d A_i z_1) \dot z_2 }{ (1- a_j^{-1} z_2 ) (1- a_k^{-1} z_2) z_1} \right] \, , \\
\langle \bs \omega, \bs n\rangle = \frac {n_1 n_2 n_3\, (J_1-J_2)(J_2-J_3)(J_3-J_1) \, G(z_1, z_2)}
{2 \det I} \left[ \frac {\dot z_1 }{ \Phi (z_1) z_2} + \frac {\dot z_2 }{\Phi (z_2) z_1} \right ], \label{<on>} \\
\Phi (z) =  (a_1^{-1}z-1) (a_2^{-1}z-1) (a_3^{-1}z-1). \nonumber
\end{gather}
Substituting them, as well as (\ref{33}), into the integrals $H, F_1$ in \eqref{H,F}, after simplifications we get
\begin{equation} \label{l,n}
\begin{aligned}
H & = (z_1-z_2) \frac {\det {\bf I} } {4 G^2(z_1,z_2)}
\left[ \frac{\Psi (z_2) }{\Phi (z_1) z_2^2} \dot z_1^2 - \frac{\Psi (z_1) }{\Phi (z_2) z_1^2} \dot z_2^2 \right]\, , \\
F_1 & = (z_1-z_2) \frac {\det {\bf I} } {4 G^2(z_1,z_2)}
\left[ \frac{\psi(z_2) }{\Phi (z_1) z_2^2} \dot z_1^2
- \frac{ \psi(z_1) }{\Phi (z_2) z_1^2} \dot z_2^2 \right]\,
\end{aligned}
\end{equation}
where
\begin{equation} \label{psis}
\Psi (z)= d \det {\bf A} z^2 - \Tr({\bf A J}) z+2 ,  \quad
\psi (z) = (4\det{\bf J}-d\, \Tr({\bf A J}))z-(\Tr {\bf J} -2d) .
\end{equation}
Next, substituting \eqref{om's}, \eqref{<on>}, and \eqref{33} into (\ref{mom}), we also obtain
\begin{equation} \label{M}
M_i = \frac {n_j n_k } 2 (J_k-J_j)
\left[ \frac{\dot z_1 }{ (1- a_j^{-1} z_1 ) (1- a_k^{-1} z_1) z_2}+
\frac{\dot z_2 }{ (1- a_j^{-1} z_2 ) (1- a_k^{-1} z_2) z_1} \right] .
\end{equation}

Now, fixing the values of the integrals by setting $H=h, F_1=f$, then solving \eqref{l,n} with respect to
$\dot z_1^2, \dot z_2^2$ and using the relation
\begin{equation} \label{G2}
\Psi (z_2) \psi (z_1)- \Psi (z_1) \psi (z_2) = \det {\bf A} \, (z_2-z_1)\, G(z_1,z_2),
\end{equation}
we get
\begin{align*}
\dot z_\alpha^2 & = - \frac { z_\beta^2 \, G(z_1,z_2) } {(z_1-z_2)^2} \frac{4 \Phi(z_\alpha) (f\Psi(z_\alpha)+ h\psi(z_\alpha)) }
{\det {\bf I} \det {\bf A} } \\
& = - \frac {z_\beta^2\, G(z_1,z_2) } {(z_1-z_2)^2} 4 (z_\alpha-a_1)(z_\alpha-a_2)(z_\alpha -a_3)\,(f\Psi(z_\alpha)+ h\psi(z_\alpha)) ,
\quad (\alpha,\beta=(1,2).
\end{align*}
After the time reparameterization
\begin{equation} \label{time}
dt = \frac {1}{\sqrt {G(z_1,z_2)} } d\tau \equiv \sqrt{\frac {\det {\bf J}}{\det {\bf I}}\, \langle {\bf I} \bs n, {\bf J}^{-1} \bs n\rangle  } \, d\tau
\end{equation}
the above relations give
\begin{gather} \label{dot_z}
\frac {d z_1}{d\tau} = \frac{z_2 \sqrt{R(z_1)} }{z_1-z_2} , \quad
\frac {d z_2}{d\tau} = \frac{z_1 \sqrt{R(z_2)} }{z_2-z_1}, \\
R(z)= -(z-a_1)(z-a_2)(z-a_3)\, (f\Psi(z)+ h\psi(z))  .  \nonumber
\end{gather}
The latter are equivalent to the following Abel--Jacobi type quadratures
\begin{gather}
\label{N4}
\frac{d z_1}{\sqrt{R (z_1)}}+\frac{d z_2}{\sqrt{R(z_2)}}=2 d\tau , \quad
\frac{z_1 \,d z_1}{\sqrt{R(z_1)}}+\frac{z_2 \, d z_2}{\sqrt{R(z_2)}}=0 ,
\end{gather}
which contain 2 holomorphic differentials on the hyperelliptic genus 2 curve
$\Gamma= \{w^2= R(z)\}$.

In view of \eqref{psis}, when $d\to 0$, the polynomial $R(z)$ becomes a degree 4 polynomial,
and \eqref{N4} reduce to the quadratures \eqref{A-J-0} for the Euler top problem, as expected.

It is interesting that, like in the integration of the original Chaplygin sphere problem presented in \cite{9},
the reparameterization factor in \eqref{time} coincides with the density \eqref{density}
of the invariant measure. In the real motion this factor never vanishes, hence the reparameterization is non-singular.

Now substituting the above formulas for $\dot z_1^2, \dot z_2^2$ into \eqref{M}, \eqref{om's} and simplifying, we express the angular momentum $M$ and the velocity $\omega$
in terms of $z_1,z_2$ and the conjugated coordinates $w_1, w_2$:
\begin{align}
M_i & = \frac{ \sqrt{ I_j I_k }\, I_i }{\sqrt{ (J_i-J_j)(J_i-J_k) }}\,
\frac { \sqrt{(a_j- z_1)(a_j- z_2)}\,\sqrt{(a_k- z_1)(a_k- z_2) } }{2 \sqrt{G(z_1,z_2) } }  \nonumber \\
 & \qquad \times \frac 1{z_1-z_2} \left[ \frac{w_1}{(z_1-a_j)(z_1-a_k)} - \frac{w_2}{(z_2-a_j)(z_2-a_k)} \right].
\label{moms_z_w} \\
\omega_i & = \frac{\sqrt{ I_j I_k} }{\sqrt{(J_i-J_j)(J_i-J_k) }}\, \frac { \sqrt{(a_j- z_1)(a_j- z_2)}\,\sqrt{(a_k- z_1)(a_k- z_2) } }{2 \sqrt{G(z_1,z_2)} } \nonumber \\
& \qquad \times \frac 1{z_1-z_2} \left[ \frac{(z_2-1/(d A_i) w_1}{(z_1-a_j)(z_1-a_k)} 
- \frac{(z_1-1/(d A_i)) w_2}{(z_2-a_j)(z_2-a_k)} \right]. \label{om_z_w}
\end{align}
Next, the projection $\langle \bs \omega, \bs n\rangle$ in \eqref{<on>} takes the form
\begin{align*}
\langle \bs \omega, \bs n\rangle & = J_1 J_2 J_3 \sqrt{ (z_1-a_1)(z_1-a_2)(z_1-a_3) \, (z_2-a_1)(z_2-a_2)(z_2-a_3) } \\
 & \qquad \times \frac 1{z_1-z_2} \left[ \frac{w_1}{ (z_1-a_1)(z_1-a_2)(z_1-a_3) }
- \frac{w_2}{ (z_2-a_1)(z_2-a_2)(z_2-a_3)} \right].
\end{align*}
Since
\begin{align*}
w_\alpha & = \sqrt{ -(z_\alpha -a_1)(z_\alpha -a_2)(z_\alpha -a_3)\, (f\Psi(z_\alpha )+ h\psi(z_\alpha )) } \\
& = \sqrt{ -(z_\alpha -a_1)(z_\alpha -a_2)(z_\alpha -a_3)\, fd \det {\bs A} (z_\alpha -c_1) (z_\alpha -c_2) } ,
\end{align*}
the latter relation also reads
\begin{align}
\langle \bs \omega, \bs n\rangle & = J_1 J_2 J_3 \sqrt{(c_1- z_1)(c_1- z_2)}\,\sqrt{(c_2- z_1)(c_2- z_2) }  \nonumber \\
& \qquad \times \frac {-1}{z_1-z_2} \left[ \frac{w_1}{(z_1-c_1)(z_1-c_2)} - \frac{w_2}{(z_2-c_1)(z_2-c_2)} \right].
\label{og_z}
\end{align}

In Section \ref{theta_sol} we shall use the above expressions to obtain explicit theta-function solutions for the
components of $n, \omega, M$ in terms of the new time $\tau$.

\section{Qualitative study of the motion and bifurcations}
For generic constants $h,f$ the polynomial $R(z)$ in \eqref{N4} has simple roots $a_i, c_1, c_2$,
and in the real motion the separating variables $z_1,z_2$ evolve between them in such a way
that $R(z_1), R(z_2)$ remain non-negative. This corresponds to a quasiperiodic motion of the sphere.

In the sequel we assume that the moments of inertia $I_1,I_2, I_3$ corresponds to a physical rigid body, i.e.,
that the triangular inequalities $I_i+I_j>I_k$ are satisfies.
For concreteness, assume also that $d<J_1<J_2<J_3$. This also implies $0<A_3<A_2<A_1$ and $0<a_1<a_2<a_3$.

As follows from the first expression in (\ref{subs1}), in the real case the factor $G(z_1, z_2)$ is always positive.
Hence, the right hand sides of \eqref{33} are positive and the coordinates $n_i$ are real and satisfy
$\langle \boldsymbol n ,\boldsymbol n \rangle =1$ if and only if
$z_1\in [a_1, a_2]$ and $z_2\in [a_2, a_3]$, like the usual spheroconical coordinates.

Next, we have

\begin{proposition} \label{c's} For the real motion, when the constants $h,f$ are positive,
and for any $d,J_i$ satisfying the above inequalities, the roots $c_1 \le c_2$ never coincide and
$$
c_1 < a_1 \quad \textup{and if $f/h=J_i$, then} \quad c_1 = \frac{2d-A_i}{d \, A_k A_j}<a_1, \quad c_2= a_i.
$$
\end{proposition}

\noindent{\it Proof.} Set $f/h=\lambda \in {\mathbb R}$. The roots $c_1, c_2$ coincide with those of
$\Psi(z)\lambda +\psi(z)$ and have the form
\begin{gather} \label{c12}
c_{1,2}=\frac { \Tr ({\bf A J}) (\lambda+d) \pm \sqrt{\mathcal D} } {d\det {\bf A}}, \\
{\mathcal D} = ( \Tr ({\bf A J}) )^2 (\lambda+d)^2-8 d \det {\bf A} (\lambda+d)+ 4d\det {\bf A}\Tr{\bf J}.
\nonumber
\end{gather}
The condition ${\cal D}=0$ gives a quadratic equation on $\lambda$, whose determinant equals
$-d \det {\bf I} \det {\bf A}^2$, always a negative number. Hence ${\cal D}>0$ and $c_1< c_2$.

Next, in view of \eqref{c12}, the condition $c_1-a_1=0$ also leads to a quadratic equation for $\lambda$,
again with always negative determinant. Then, evaluating $c_1-a_1$ for one value of $\lambda$, we find $c_1 < a_1$ 
for any $\lambda\in {\mathbb R}$.

Finally, setting in \eqref{c12} $\lambda=J_i$ and simplifying, we obtain the indicated above expressions for $c_1, c_2$.
\medskip

Combining the statement of Proposition \ref{c's} with the permitted positions of $z_1, z_2$, we conclude that,
depending on value of $c_2$,
\begin{equation} \label{ccc}
z_1 \in [a_1, c_2], \;  z_2 \in [a_2, a_3], \quad \mbox{or } \quad z_1 \in [a_1, a_2], \;  z_2 \in [c_2, a_3].
\end{equation}
Then, in view of \eqref{33}, the vector $\boldsymbol n$ always fills a ring ${\mathcal R}$ on
the unit sphere $S^2=\{\langle x, x\rangle =1 \}$ between the lines of its intersection with the cone
$$
\sum_{i=1}^3 \frac{J_i-d}{J_i} \frac {x_i^2}{a_i-c_2}=0 .
$$

\paragraph{Periodic solutions with bifurcations.} As follows from Proposition \ref{c's}, the only periodic solutions
with bifurcations can occur when the root $c_2$ coincides with $a_1,a_2$ or $a_3$\footnote{There is another type of
periodic solutions corresponding to periodic windings of the 2-dimensional tori. However, the latter are not related
to bifurcations and we do not consider them here.}. This happens under the initial conditions
$$
\omega_i=\omega_j =0, \quad n_k=0, \qquad (i,j,k)=(1,2,3),
$$
when the sphere performs a periodic circular motion with $ n_k \equiv 0$ and
one has $\langle \omega,n\rangle \equiv 0$, $H=J_k \omega_k$, $F_1=J_k^2 \omega_k^2$, which yields $f/h=J_k$.
Then, in view of the above proposition, $c_2=a_k$ and the polynomial $R(z)$ in \eqref{dot_z} has the double root $a_k$, as expected.

When $\lambda=f/h$ leaves the interval $[J_1,J_3]$, the root $c_2$ goes beyond of $[a_1, a_3]$.
Then for $R(z_1), R(z_2)$ to be both positive, one of $z_i$ must violate the condition \eqref{ccc}.
This implies that in the real case the quotient $f/h$ belongs to $[J_1, J_3]$, and
the bifurcation diagram on the plane $(h,f)$ consists only of 3 rays $f/h=J_1, J_2, J_3$.

Note that, according to the results of \cite{9}, a similar situation takes place for the Chaplygin sphere on the horizontal plane.

\paragraph{The motion of the contact point on the fixed sphere.} As mentioned above, in the generic case
 with $F_2=0$ the contact point on the moving sphere given by the vector ${\boldsymbol n}$
belongs to the ring $\cal R$ on $S^2$.

 Then the following natural question arises: {\it does the contact point on the {\it fixed} sphere also belongs to a ring
 or it cover the whole sphere} ?
(Recall (\cite{9}) that in the case of the Chaplygin sphere on a horizontal plane the contact point on the plane moves
inside a strip, whose axis is orthogonal to the horizontal momentum vector.)

To study the above problem we assume, without loss of generality, that the radius $a$ of the fixed sphere is 1. 
Then the contact point on this sphere is given by the unit vector ${\boldsymbol n}$ as viewed {\it in space}. 

To describe the spatial evolution of $\bs n$,
introduce a fixed orthogonal frame $O\xi \eta \zeta$ with the center $O$ in the center of the fixed sphere and 
the ``vertical'' 
axis $O\zeta$ directed along the fixed momentum vector $\bs M$. Then, in view of \eqref{M}, the projection of
${\boldsymbol n}$ on $O\zeta$ can be written in form
\begin{align*}
n_\zeta & = \frac {1}{|\bs M |} \langle {\boldsymbol n}, \bs M \rangle
= \frac {1}{\sqrt{f} } a_1 a_2 a_3 \,
(J_1-J_2)(J_2-J_3)(J_3-J_1) \, n_1 n_2 n_3  \\
& \quad \times \left[ \frac {z_1 \, \dot z_1}{(z_1-a_1)(z_1-a_2 )(z_1-a_3) z_2 } +
\frac {z_2 \, \dot z_2}{(z_2-a_1)(z_1-a_2 )(z_1-a_3) z_1 } \right ],
\end{align*}
which, following \eqref{33} and the expressions for ${\dot z}^2_1, {\dot z}^2_2$, after a simplification, reads
\begin{align}
n_\zeta & = \frac {\sqrt{-1}}{\sqrt{f} }
J_1 J_2 J_3 \, 
\sqrt{(c_1- z_1)(c_1- z_2)}\,\sqrt{(c_2- z_1)(c_2- z_2) }  \nonumber \\
& \quad \times \frac {1}{G(z_1, z_2)} \frac {-1}{z_1-z_2}
\left[ \frac{z_2 w_1}{(z_1-c_1)(z_1-c_2)} - \frac{z_1 w_2}{(z_2-c_1)(z_2-c_2)} \right].
\label{og_n}
\end{align}

It follows that the right hand side of \eqref{og_n} is a quasiperiodic function of time. One can show that
under the conditions \eqref{ccc} and $c_1<a_1<a_2<a_3$, the function $n_\zeta(z_1,w_1,z_2,w_2)$
is real and, regardless to signs of the roots $w_\alpha =\sqrt{R(z_\alpha )}$, the function $|n_\zeta|$
has the absolute maximum in one of the vertices of the quadrangle ${\cal Q} \subset (z_1,z_2)={\mathbb R}^2$ defined by \eqref{ccc}.  In two other vertices of ${\cal Q}$ this function is zero.

Calculating $|n_\zeta|$ in the vertices of ${\cal Q}$, we find that for $c_2\ne a_i$ its maximum is strictly less than 1. 
It follows that the trajectory $\bs n (t)$ on the fixed sphere lies between the
``horizontal'' planes $\zeta=\pm\nu$, $\nu < 1$.

\medskip

To describe the trajectory $\bs n (t)$ on the fixed sphere in the ``longitudinal'' direction, apart from 
the fixed momentum vector $\bs M$ it is good to know another fixed vector which can be expressed in terms of $\omega, \bs n$. However, it seems that such a vector does not exist, and for this reason we introduce the longitude angle $\psi$ between the axis $O\xi$ and the vertical plane spanned by $\bs M$ and $\bs n$. Introduce also the
the longitudinal unit vector $\bs u= M\times \bs n / |\bs M\times \bs n| $. Then we find
$$
\dot \psi = \frac {|\bs M|}{ |\bs M\times \bs n| } \left\langle \bs u , \frac d{dt} {\boldsymbol n} \right\rangle = 
|\bs M| \frac { \left\langle \bs M\times \bs n , \frac d{dt} {\boldsymbol n} \right \rangle }
{ \langle  \bs M\times \bs n , \bs M\times \bs n \rangle } ,
$$
where $\frac d{dt} {\boldsymbol n}$ is the absolute derivative of ${\boldsymbol n}$ expressed in
the coordinates of the moving frame. 
In view of the second vector equation in (2.1) with $k=-1$, 
$$
\frac d{dt} {\boldsymbol n} = \dot {\boldsymbol n} + \omega \times {\boldsymbol n} = 2 \omega \times {\boldsymbol n}.
$$
Hence, we get
\begin{equation}
\label{dotpsi}
\dot \psi = |\bs M| \frac { \left\langle \bs M\times \bs n , 2 \omega \times {\boldsymbol n} \right\rangle }
{ \langle  \bs M\times \bs n , \bs M\times \bs n \rangle }. 
\end{equation}
Next, using the expressions \eqref{om's}, \eqref{M}, we obtain
 \begin{gather*}
 2 (\omega \times \bs n)_i = \frac{n_i (J_i-d)}{G} \left[ \frac{d A_j A_k z_2+A_i-2d}{a_i^{-1}z_1-1}\dot z_1 +
 \frac{d A_j A_k z_1 +A_i-2d}{a_i^{-1}z_2-1}\dot z_2 \right] , \\
 (\bs M \times \bs n)_i = \frac{n_i (J_i-d)}{2 G} \left[ \frac{ (2J_j J_k-d A_i) z_2 - 1 }{(a_i^{-1}z_1-1) z_2} \dot z_1 + \frac{(2J_j J_k-d A_i ) z_1 - 1 }{(a_i^{-1}z_2-1)z_1}\dot z_2 \right] , \\
(i,j,k) =(1,2,3) .
 \end{gather*}
Substituting these formulas into \eqref{dotpsi} and expressing the derivatives $\dot z_\alpha$ in terms of 
$z_1,w_1,z_2,w_2$, one finds the derivative $\dot\psi$ as a symmetric function of $(z_1,w_1)$ and $(z_2,w_2)$, that is, as a quasiperiodic function of $t$. Its integration yields $\psi (t)$, which, together with \eqref{og_n}, provides
a complete description of the contact point on the fixed sphere.

\section{A special case of periodic motion.} Apart from the particular case of the motion with $F_2=0$,
there is another special case, when this integral takes the maximal value, that is, when ${\bf A}\bs n$
is parallel to the momentum
vector $\bs M$. In this case $\bs M=h\, {\bf A}\bs n$, $h=$const. In view of \eqref{mom}, this implies
\begin{equation} \label{M_n}
{\bf J}\bs\omega -d \langle \bs\omega, \bs n\rangle = h\, {\bf A}\bs n \quad \mbox{and} \quad
\bs \omega= h {\bf J}^{-1} \left( {\bf A} \bs n + \frac{d}{\rho^2}\langle {\bf A} \bs n, {\bf J}^{-1} \bs n\rangle \right),
\end{equation}
$\rho$ being the same as in (\ref{density}).

Substituting the expression for $\omega$ into the second equation in \eqref{1} and simplifying, we get
the following closed system for $n$:
\begin{equation} \label{Euler}
\bs  n = h \frac{J_1+J_2+J_3-2d}{F}\, (\bs n \times {\bf J}^{-1} \bs n) \, .
\end{equation}
It has two independent integrals $\langle \bs n, \bs  n\rangle$ and $\langle \bs n, {\bf J}^{-1} \bs n\rangle$
or $\langle {\bf A} \bs n, {\bf A} \bs n\rangle$,
which implies that the factor $F$ is constant on the trajectories and that the system has the form of the Euler top
equations.
As a result, in the general case the components of $n$ and $\omega$ are expressed in terms of elliptic functions of
the original time $t$ and their evolution is periodic.

This situation is similar to that of the special case of the motion of the Chaplygin sphere on a horizontal plane, when
the momentum vector $\bs M$ is vertical, and when the solutions are elliptic in the original time.


\section{Theta-function solutions in the case $F_2=0$.} \label{theta_sol}
In order to find explicit solutions for the components of $\omega, \bs M, \bs n$, and other variables, we first
remind some necessary basic facts on the Jacobi inversion problem and its solution.

\begin{figure}[h,t]
\begin{center}
\includegraphics[height=0.25\textwidth, width=0.75\textwidth]{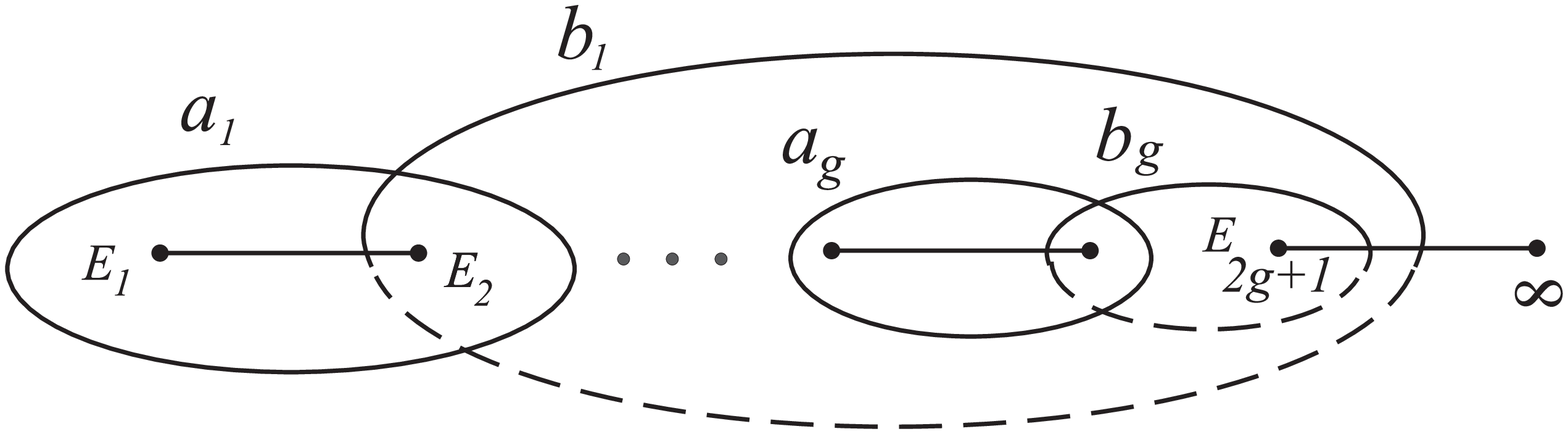}
\end{center} \caption{\footnotesize{A canonical basis of cycles on the hyperelliptic curve represented as 2-fold covering
of the complex plane $\lambda$. The parts on the cycles on the lower sheet are shown by dashed lines.} }\label{1.fig}
\end{figure}

\paragraph{Solving the Jacobi inversion problem by means of Wurzelfunktionen.}
Consider an {\it odd-order} genus $g$ hyperelliptic Riemann surface $\Gamma $ obtained from the affine curve
$$
\{\mu^{2}=R(\lambda )\}, \qquad
R(\lambda )=(\lambda -E_{1})\cdots(\lambda -E_{2g+1})\} ,
$$
by adding one infinite point $\infty$. Let us choose a canonical basis of cycles
${\mathfrak a}_{1},\ldots, {\mathfrak a}_{g}$,
${\mathfrak b}_{1},\ldots, {\mathfrak b}_{g}$ on $\Gamma$ such that
$$
{ \mathfrak a}_{i}\circ {\mathfrak a}_{j}={\mathfrak b}_{i}\circ {\mathfrak b}_{j}=0,
\quad  {\mathfrak a}_{i}\circ {\mathfrak b}_{j}=\delta_{ij}, \qquad
i,j=1,\ldots,g,
$$
where $\gamma_{1}\circ \gamma_{2}$ denotes the intersection index of the cycles
$\gamma_{1},\gamma_{2}$
(For real branch points see an example in Figure \ref{1.fig}). Next, let $\bar{\omega}_{1},\dots,\bar{\omega}_{g}$
be the conjugated basis of normalized holomorphic differentials on $\Gamma$ such that
$$
\oint_{{\mathfrak a}_j}{\bar\omega}_{i}=2\pi \jmath \,\delta_{ij},
\qquad \jmath =\sqrt{-1}.
$$
The $g\times g$ matrix of $b$-periods $B_{ij}=\oint_{{\mathfrak b}_j}\bar\omega_{i}$
is symmetric and has a negative definite real part. Consider the period lattice
$\Lambda^0=\{2\pi \jmath{\mathbb Z}^{g}+B{\mathbb Z}^{g}\}$ of rank $2g$
 in ${\mathbb C}^{g}=(Z_1,\dots, Z_g)$.
The complex torus Jac$(\Gamma )={\mathbb C}^{g}/\Lambda^0$ is called the Jacobi
variety ({\it Jacobian}) of the curve $\Gamma $. For a fixed point $P_0$ the Abel map
$$
{\cal A}: \Gamma \mapsto {\rm {Jac}}(\Gamma), \quad {\cal A}(P)
= \int_{P_0}^P (\bar{\omega}_{1},\dots,\bar{\omega}_{g})^{T}
$$
describes a natural embedding of the curve into its Jacobian.

Now consider a generic divisor of points
$P_{1}=(\lambda _{1},\mu_{1}),\ldots, P_{g}=(\lambda _{g},\mu_{g})$
on it, and the Abel--Jacobi mapping with a basepoint $P_0$
\begin{gather}
\label{1.16}
\int^{P_{1}}_{P_{0}}\bar{\omega}+\cdots+\int^{P_{g}}_{P_{0}}\bar{\omega}=Z, \\
\bar{\omega}= (\bar{\omega}_{1},\dots,\bar{\omega}_{g})^{T} , \quad
Z=(Z_{1},\ldots,Z_{g})^{T}\in {\mathbb C}^{g} .\nonumber
\end{gather}
Under the mapping, symmetric functions of the coordinates of the points
$P_{1},\ldots,P_{g}$ are $2g$-fold periodic functions of the complex variables
$Z_{1},\ldots,Z_{g}$ with the above period lattice $\Lambda^{0}$ (Abelian functions).

Explicit expressions of such functions can be obtained
by means of theta-functions on the universal covering  ${\mathbb C}^g=(Z_{1},\ldots,Z_{g})$
of the complex torus.
Recall that customary Riemann's theta-function $\theta(Z|B)$ associated with the
Riemann matrix $B$ is defined by the series\footnote{The expression for $\theta(Z)$ we use
here is different from that chosen in a series of books on theta-functions by
multiplication of $Z$ by a constant factor.}
\begin{gather}
\theta (Z|B)= \sum_{M\in {\mathbb Z}^{g}}\exp \langle \langle BM,M\rangle +\langle M,Z\rangle ),  \label{1.theta-def}  \\
\langle M,Z\rangle =\sum^{g}_{i=1}M_{i}Z_{i}, \quad
\langle BM,M\rangle =\sum^{g}_{i,j=1}B_{ij}M_{i}M_{j} .\nonumber
\end{gather}

Equation $\theta(Z|B)=0$ defines a codimension one
subvariety $\Theta\in \mbox{Jac} (\Gamma)$ (for $g>2$ with singularities)
called {\it theta-divisor}.


We shall also use {\it theta-functions with characteristics}
$$
\alpha =(\alpha_{1},\ldots,\alpha_{g}), \quad
\beta =(\beta_{1},\ldots,\beta_{g}), \qquad \alpha_j,\beta_j \in{\mathbb R},
$$
which are obtained from $\theta (Z|B)$ by shifting the argument $Z$ and multiplying by
an exponent\footnote{Here and below we omit $B$ in the theta-functional notation.}:
$$
\theta\! \left[{ \alpha \atop \beta}\right]\! (Z)
\equiv \theta \! \left[{\alpha_{1}\, \cdots \, \alpha_{g} \atop
                       \beta_{1} \, \cdots \, \beta_{g}} \right]\! (Z)
=\exp \{\langle B\alpha ,\alpha \rangle /2+\langle Z+2\pi  \jmath\beta,\alpha \rangle \}\,
\theta (Z+2\pi \jmath\beta+B\alpha) .
$$
All these functions enjoy the quadiperiodic property
\begin{eqnarray}
\theta\! \left[ { \alpha \atop \beta } \right ] \! (Z+2\pi \jmath K+BM)
=\exp (2\pi \jmath\epsilon)\exp \{-\langle BM,M\rangle /2-\langle M,Z\rangle \}
\theta\! \left[ { \alpha \atop \beta } \right ]  \!(Z) , \label{1.5} \\
\epsilon =\langle \alpha ,K\rangle -\langle \beta ,M\rangle  , \nonumber
\end{eqnarray}

\medskip

Now for a generic divisor
$P_{1}=(\lambda _{1},\mu_{1}),\ldots, P_{g}=(\lambda _{g},\mu_{g})$
on $\Gamma $, introduce the polynomial $U(\lambda ,s)=(s-\lambda _{1})\cdots(s-\lambda _{g})$,
$\lambda\in {\mathbb C}$. It is known (see e.g., \cite{Baker, BEL}) that given a generic constant $C\ne E_i$, then
under the Abel mapping (\ref{1.16}) with $P_0=\infty$ the following relations hold
\begin{gather} \label{ab0}
U(\lambda, C) \equiv (C -\lambda_1)\cdots(C-\lambda_g) = 
\varkappa \frac{ \theta [\Delta](Z-q) \theta [\Delta](Z+q) }{ \theta^2 [\Delta](Z)}, \\
   q= {\cal A} (C, \sqrt{R(C)} ) = \int_{\infty}^{(C, \sqrt{R(C)} )} (\bar{\omega}_{1},\dots,\bar{\omega}_{g})^{T},
\nonumber   
\end{gather}
where  $\varkappa$ is a constant depending on the periods of $\Gamma $ only. 

These relation can be generalized in different ways as follows.

\begin{theorem} \label{Wurzel} \textup{(see, e.g., \cite{Baker, BEL, Acta_bill}).}
Under the Abel mapping \textup{(\ref{1.16})} with $P_0=\infty$ the following relations hold
\begin{align}
\sqrt{U(\lambda,E_i)}\equiv \sqrt{(E_i-\lambda_1)\cdots(E_i-\lambda_g)}
&= k_{i}\frac{\theta [\Delta+\eta_{i}](Z) } {\theta[\Delta](Z)},     \label{ab2.1} \\
\sum^{g}_{k=1} \frac{\mu_k} {\prod_{l\ne k}(\lambda_k-\lambda_l)}\,
\frac {\sqrt{U(\lambda,E_i)}\, \sqrt{U(\lambda,E_j)}} {(E_{i}-\lambda_{k})(E_{j}-\lambda_{k})}
& = k_{ij} \frac{\theta[\Delta+\eta_{ij}](Z) } { \theta [\Delta](Z)}, \label{ab2.2} \\
i,j=1,\dots,2g-1, \quad i & \neq j , \nonumber
\end{align}
where  $k_{i}, k_{ij}$ are certain constants depending on the periods of
$\Gamma $ only, and
$$
\Delta=\begin{pmatrix} \Delta' \\ \Delta'' \end{pmatrix}, \quad
\eta_i=\begin{pmatrix} \eta_i' \\ \eta''_i \end{pmatrix}, \qquad
\Delta' ,\Delta'',   \eta_i' , \eta''_i \in \frac 12 {\mathbb Z}^g/{\mathbb Z}^g
$$
are half-integer theta-characteristics such that
\begin{align}
2\pi \jmath \, \eta_i''+B\eta_i' & = \int^{(E_i,0)}_{\infty}\bar{\omega}
\quad ({\rm mod} \;\Lambda),   \label{eta} \\
 2\pi i\Delta''+B\Delta' & ={\cal K} \; ({\rm mod}\; \Lambda), \quad \mbox{and} \quad
\eta_{ij}=\eta_{i}+\eta_{j} \quad ({\rm mod} \;{\mathbb Z}^{2g}) , \nonumber
\end{align}
${\cal K}\in {\mathbb C}^g$ being the vector of the Riemann constants.
\end{theorem}

Apparently, relations (\ref{ab2.2}) were first obtained in the explicit
form by K\"onigsberger (\cite{Konigs}). Earlier, expressions (\ref{ab2.1}) had been
considered by K.Weierstrass as generalizations of the Jacobi
elliptic functions ${\rm sn}(Z), {\rm cn}(Z)$, and dn$(Z)$.
This set of remarkable relations
between roots of certain functions on symmetric products of
hyperelliptic curves and quotients of theta-functions with half-integer
characteristics is historically referred to as {\it Wurzelfunktionen} (root functions).
\medskip

One can show (see, e.g., \cite{Fay, Baker}) that
for the chosen canonical basis of cycles
${\mathfrak a}_{1},\ldots,{\mathfrak a}_{g}$, ${\mathfrak b}_{1},\ldots,{\mathfrak b}_{g}$ on
$\Gamma$, 
\begin{equation} \label{deltas_g}
\Delta '=(1/2,\ldots,1/2)^{T}, \quad \Delta''=(g/2,(g-1)/2,\ldots,1,1/2)^{T} \quad
({\rm mod}\; 1).
\end{equation}

In the case $g=2$, all the functions in (\ref{ab2.1}), (\ref{ab2.2}) are
single-valued on the 16-fold covering ${\mathbb T}^{2}\rightarrow $Jac$(\Gamma )$
with each of the four periods of $\Lambda_{0}$ doubled, so that
${\mathbb T}^{2}$ and Jac$(\Gamma )$ are transformed to each
other by the change $Z\rightarrow 2Z.$
In view of (\ref{eta}), (\ref{deltas_g}), one has
\begin{align}
\Delta & =\begin{pmatrix} 1/2&1/2\cr0&1/2 \end{pmatrix} ,\quad
 \Delta +\eta _{1}=\begin{pmatrix} 0&1/2\cr0&1/2 \end{pmatrix}, \quad
\Delta +\eta _{2}=\begin{pmatrix} 0&1/2\cr1/2&1/2 \end{pmatrix}, \nonumber \\
\Delta +\eta _{3}& =\begin{pmatrix} 1/2&0\cr1/2&1/2\end{pmatrix}, \quad
\Delta +\eta _{4}=\begin{pmatrix} 1/2&0\cr1/2&0\end{pmatrix}, \quad
 \Delta +\eta _{5}=\begin{pmatrix} 1/2&1/2\cr1/2&0 \end{pmatrix} . \label{deltas}
\end{align}

We shall also need the following modification of the K\"onigsberger formula \eqref{ab2.2}, which, for our
convenience, we adopt for the case $g=2$.

\begin{theorem} \label{Wurzel_2} Let $C\in {\mathbb C}$ be a constant that does not coicide with $E_i$.
Then under the Abel mapping \textup{(\ref{1.16})} with $g=2$ and $P_0=\infty$: 
\begin{gather}
 \frac{1}{\lambda_1-\lambda_2} \left[ \frac{\mu_1}{ (E_{i}-\lambda_{1})(E_{j}-\lambda_{1})(C-\lambda_1) } -
\frac{\mu_2}{ (E_{i}-\lambda_{2})(E_{j}-\lambda_{2})(C-\lambda_2) }  \right] \nonumber \\
 = \hat k_{ij} \frac{\theta^2 [\Delta](Z)\; \theta[\Delta+\eta_{ij}](Z-q)\; \theta[\Delta+\eta_{ij}](Z+q) }
  {\theta [\Delta+ \eta_i ](Z ) \; \theta [\Delta+ \eta_j](Z) \; \theta [\Delta] (Z -q) \,  \theta [\Delta] (Z+q) },
 \label{block} \\
i,j=1,\dots,2g-1, \quad i  \neq j , \nonumber
\end{gather}
where $\hat k_{ij}$ are constants depending on the periods of $\Gamma$ only and, as in \eqref{ab0}, 
$q={\cal A}(C,\sqrt{R(C)})$.
\end{theorem}

\noindent{\it Proof.} Let us fix the point $P_2=(\lambda_2, \mu_2)$ in a {\it generic}
position on the curve $\Gamma$ and consider the following meromorphic function on this curve
$$
f(P) = \frac{\mu + \mu_2 \frac{ (\lambda-E_i ) (\lambda-E_j ) (\lambda-C )  ) }
{ (\lambda_2-E_i ) (\lambda_2-E_j ) (\lambda_2-C ) }  }{(\lambda-\lambda_2) (\lambda-E_i ) (\lambda-E_j ) (\lambda-C ) } .
$$
Due to the order of poles and zeros of $z,w$, and $\lambda-E_i$, $\lambda-C$ on $\Gamma$, for any generic $P_2$, the function $f(P)$
has simple poles at $P=P_2, E_i, E_j, Q_-, Q_+$ and does not have a pole neither at $\iota P_2=(\lambda_2,-\mu_2)$, nor at any other point on $\Gamma$.  Next, $f(P)$ has a double zero at $\infty$. Then, using the description of zeros of
$\theta(Z)$, $\theta[\Delta+ \eta_i ](Z), \theta [\Delta+\eta_{ij}](Z)$ one can show that up to a constant factor, $f(P)=\bar f(P)$ with
\begin{align*}
\bar f(P) & = \frac{ \theta^2 [\Delta] ( {\cal A}(P) -{\cal A}(P_2) )}
{ \theta [\Delta+ \eta_i ]( {\cal A}(P)-{\cal A}(P_2) ) \, \theta [\Delta+ \eta_j]({\cal A}(P)-{\cal A}(P_2) )}  \\
&\quad \times \frac{\theta [\Delta+ \eta_{ij} ] ({\cal A}(P) -q -{\cal A}(P_2) )\,
\theta [\Delta+\eta_{ij}] ({\cal A}(P) +q -{\cal A}(P_2)) }{ \theta [\Delta] ({\cal A}(P) -q -{\cal A}(P_2) ) \,  \theta [\Delta] ({\cal A}(P) +q -{\cal A}(P_2) ) } .
\end{align*}
Note that due to the quasiperiodic property of the theta-functions with characteristics, 
$\bar f(P)$ is a meromorphic function on $G$.

Now setting $P=\iota P_1=(\lambda_1,-\mu_1)$, the function $-f(P)$ transforms to the left hand side of \eqref{block}, and
the argument ${\cal A}(P) -{\cal A}(P_2)$ becomes $-Z$. Hence $\bar f(P)$ transforms to
the right hand side of \eqref{block}, which proves the theorem.
\medskip

Combining Theorem \ref{Wurzel_2} and formula (\ref{ab0}), we obtain the following useful corollary.

\begin{proposition} 
Under the Abel mapping \textup{(\ref{1.16})} with $g=2$ and $P_0=\infty$, 
\begin{gather}
 \frac{1}{\lambda_1-\lambda_2} \left[ \frac{(C-\lambda_2) \mu_1}{ (E_{i}-\lambda_{1})(E_{j}-\lambda_{1}) } -
\frac{(C-\lambda_1) \mu_2}{ (E_{i}-\lambda_{2})(E_{j}-\lambda_{2}) }  \right] \nonumber \\
 = {\rm const}_{ij} \frac{ \theta[\Delta+\eta_{ij}](Z-q)\; \theta[\Delta+\eta_{ij}](Z+q) }
  {\theta [\Delta+ \eta_i ](Z ) \, \theta [\Delta+ \eta_j](Z) },
 \label{small_block} \\
q={\cal A}(C,\sqrt{R(C)}), \qquad i,j=1,\dots,2g-1, \quad i  \neq j , \quad C\ne E_i. \nonumber
\end{gather}
\end{proposition}

\noindent{\it Proof.} Indeed, the left hand side of \eqref{small_block} is obtained from that of 
\eqref{block} by multiplication by $(C-\lambda_1)(C-\lambda_2)$, 
whose theta-function expression is given by formula (\ref{ab0}). Then the product of right hand sides 
of (\ref{block}) and (\ref{ab0}) gives \eqref{small_block}.  

\paragraph{Explicit theta-function solutions.} Now let $\Gamma $ be the genus 2 curve
$$
\{w^2= R(z)\}, \qquad R(z)= -(z-a_1)(z-a_2)(z-a_3)\, (z-c_1)(z-c_2)\,,
$$
$c_1 < c_2$ being the roots of $f\Psi(z)+ h\psi(z)$. Thus we identify (without order)
$$
\{E_1, \dots, E_5 \}= \{a_1, a_2, a_3, c_1, c_2\},
$$
and denote the corresponding half-integer characteristic $\eta_i$ by $\eta_{a_i}$ and $\eta_{c_\alpha}$.

Next, choose the canonical basis of cycles as depicted in Fig. \ref{1.fig} and
calculate the $2\times 2$ period matrix
$$
A_{ij}=\oint_{{\mathfrak a}_j}\varpi_{i}, \qquad \varpi_1= \frac{dz}{\sqrt{R(z)}}, \quad
\varpi_2 = \frac{z\, dz}{\sqrt{R(z)}} .
$$
Then the normalized holomorphic differentials on $\Gamma$ are
$$
\bar{\omega}_{k}=\sum_{j=1}^2 C_{kj} \frac{z^{j-1} d z}{\sqrt{R(z)}}, \qquad  C=A^{-1},
$$
and the quadratures (\ref{N4}) give
\begin{gather}
\int_\infty^{ (z_1,w_1) } \bar\omega_1 + \int_\infty^{(z_2,w_2)} \bar\omega_1 = Z_1 , \quad
\int_\infty^{ (z_1,w_1) } \bar\omega_2 + \int_\infty^{(z_2,w_2)} \bar\omega_2 = Z_2 , \label{int_Abel} \\
 Z_1 = 2 C_{11} \tau + Z_{10}, \quad  Z_2 = 2C_{21} \tau + Z_{20},  \label{Z-tau}
\end{gather}
$Z_{10}, Z_{20}$ being constant phases.

Now, comparing the last fraction in \eqref{33} with the expression (\ref{ab2.1}) in Theorem \ref{Wurzel}, we
find
\begin{equation} \label{S_i}
S_i \equiv \frac{\sqrt{ (a_i- z_1)(a_i- z_2 )} }{\sqrt{(a_i-a_j)(a_i- a_k)}} = k_i
\frac{ \theta[\Delta+ \eta_{a_i}](Z) }{ \theta[\Delta](Z) } ,  \qquad (i,j,k)=(1,2,3),
\end{equation}
where $Z=(Z_1,Z_2)$ and the components of $Z$ depend on $\tau$ according to \eqref{Z-tau}. 

To calculate $k_i$, we set here $z_1=a_j, z_2= a_k$. 
Then, in view of (\ref{int_Abel}), the definition of $\theta$ with characteristics and
the quasiperiodic property \eqref{1.5},
$$
1= k_i \frac{ \theta[\Delta+ \eta_{a_i} ]( {\cal A}(a_j)+ {\cal A}(a_k) ) }
{ \theta[\Delta]( {\cal A}(a_j)+ {\cal A}(a_k) ) }
= k_i \frac{ \theta[\Delta]( 0) }
{ \theta[\Delta+  \eta_{a_i} ](0)}, \quad \mbox{that is,} \quad 
k_i = \frac { \theta[\Delta+\eta_{a_i} ](0)}{\theta[\Delta](0)} .    
$$

In order to express in theta-functions the factor $G(z_1, z_2)$ 
given by \eqref{G}, we fist note that
it does not split into a product of linear functions in $z_1$ and $z_2$,
hence one cannot use the formula (\ref{ab0}).

On the other hand, from the condition
$n_1^2+n_2^2+n_3^2 \equiv 1$ we find that
$$
 G= \frac{\det {\bf I}}{\det {\bf J}} \left ( \frac{J_1}{J_1-d} S_1^2+ \frac{J_2}{J_2-d} S_2^2+  \frac{J_3}{J_3-d} S_1^2 \right),
$$
which, in view of \eqref{S_i}, gives
\begin{gather} \label{G_theta}
  \sqrt{G(z_1, z_2)} = \sqrt{ \frac{\det {\bf I}}{\det {\bf J}} }\, \frac {\sqrt{\Sigma(Z)}}{ \theta [\Delta](Z) }, \\ 
  \Sigma (Z)= \frac{J_1}{I_1} \theta^2[\Delta+ \eta_{a_1}](Z) + \frac{J_2}{I_2} \theta^2[\Delta+ \eta_{a_2}](Z)
 + \frac{J_3}{I_3} \theta^2 [\Delta+ \eta_{a_3}](Z) . \nonumber
\end{gather}
As a local singularity analysis shows, in general the function $\Sigma(Z)$ has zeros of first order, 
hence it cannot be a full square of another theta-function expression. 


Now, using theta-function expressions \eqref{S_i}, \eqref{G_theta} in formulas \eqref{33}, \eqref{moms_z_w}, 
\eqref{om_z_w} and applying also the Wurzelfunktionen (\ref{ab2.2}), \eqref{small_block} with $C=1/(dA_i)$,
we arrive at the following theorem.

\begin{theorem} \label{main} The generic theta-function solutions for the Chaplygin sphere-sphere problem 
in the case $F_2=0$ have the form
\begin{align}
n_i (\tau) & = \kappa_i \frac {\theta[\Delta+ \eta_{a_i}](Z)} {\sqrt{\Sigma(Z)} }, \\ 
M_i (\tau) & = \nu_i \frac {\theta[\Delta+ \eta_{a_j}+\eta_{a_k} ](Z)} {\sqrt{\Sigma(Z)} }, \\
\omega_i(\tau) &
= \varepsilon_i \frac { \theta[\Delta+ \eta_{a_j}+\eta_{a_k} ](Z-q_i)\, \theta[\Delta+ \eta_{a_j}+\eta_{a_k} ](Z+q_i)}
{\theta[\Delta](Z) \cdot \sqrt{\Sigma (Z)} }, \\
& \qquad q_i  = {\cal A}(1/(dA_i), \sqrt{R( 1/(dA_i))}), \quad \kappa_i, \nu_i, \varepsilon_i = \textup{const}, 
\quad  (i,j,k)=(1,2,3), \nonumber
\end{align}
where the characteristics are given in \textup{ (\ref{deltas})} and $Z_1, Z_2$ depend linearly on $\tau$
as described in \eqref{Z-tau}.
\end{theorem}

We do not give explicit expressions for the constants $\kappa_i, \nu_i, \varepsilon_i$ here.  

Note that due to presence of the square root, 
the variables $n_i, M_i, \omega_i$ are not meromorphic functions of $Z_1,Z_2$ and therefore, of the new time $\tau$. This
stays in contrast with the solutions of the classical Chaplygin sphere problem, which, 
after a similar time reparameterization, become meromorphic (see \cite{Dust, Ch_sol}). 


Next, comparing the expression \eqref{og_z} with the Wurzelfunktion (\ref{ab2.2}) and \eqref{og_n} with 
\eqref{small_block}, assuming $C=0$, we also find 
\begin{align}
\langle \bs \omega, \bs n\rangle & = \upsilon \frac { \theta[\Delta+ \eta_{c_1}+\eta_{c_2} ](Z)} {\theta[\Delta](Z) }, \\
n_\zeta & = \varrho \frac { \theta[\Delta+ \eta_{c_1}+\eta_{c_2} ](Z-\hat q)\, \theta[\Delta+ \eta_{c_1}+\eta_{c_2} ](Z+\hat q)}
{\theta[\Delta](Z) \; \sqrt{\Sigma (Z)} }, \\
& \qquad \hat q = {\cal A} (0,\sqrt{R(0)}), \qquad \upsilon, \varrho=\textup{const}. \nonumber
\end{align}
The second formula, together with \eqref{Z-tau}, describes the altitude of the contact point on the fixed sphere as a function of $\tau$. 

Finally, given the expression \eqref{G_theta} for the factor $G$, the original time $t$ can be found as a function 
of $\tau$ by integrating the quadrature (\ref{time}).

\section*{Appendix. Separation of variables via reduction to a Hamiltonian system}
As mentioned above, the substitution \eqref{33} is quite
non-trivial and can hardly be guessed a priori. Below we describe how one can obtain it in a systematic way.

\subsection*{A1. Reduction to a Hamiltonian system on $S^2$}
First introduce the spheroconical coordinates $u,\,v$ on the Poisson sphere
$\langle {\boldsymbol n}, {\boldsymbol n} \rangle=1$:
\begin{equation}
\label{333}
n_i^2=\frac{(J_i-u)(J_i-v)}{(J_i-J_j)(J_i-J_k)},\quad i\neq j\neq k\neq
i,\quad J_i=I_i+d.
\end{equation}
Then, under the substitution \eqref{3}, the equations \eqref{1} with $k=-1$ give rise to the following Chaplygin-type system on $S^2$
\begin{gather}
\label{4}
\frac{d}{dt}\frac{\partial T}{\partial\dot u}-\frac{\partial T}{\partial u} =\dot u\Phi, \quad
\frac{d}{dt}\frac{\partial T}{\partial\dot v}-\frac{\partial T}{\partial v} =-\dot v\Phi\, , \\
T=\frac{1}{2}(b_{uu}{\dot u}^2+b_{uv}{\dot u}{\dot v}+b_{vv}{\dot v}^2), \quad
\Phi=(a_{u}{\dot u}+a_{v}{\dot v}), \nonumber
\end{gather}
where $T$ is the energy integral \eqref{2} expressed in the spheroconical coordinates under the condition
$F_2=0$ and $\Phi$ is linear homogeneous in $\dot u, \dot v$. Explicit expressions for the coefficients
of $T,\,\Phi$ are quite tedious, so we do not give them here.

Introducing the momenta $P_u=\displaystyle\frac{\partial T}{\partial\dot u}$, $P_v=\displaystyle\frac{\partial T}{\partial\dot v}$,
this system can be transformed to a Hamiltonian form with extra terms, which possesses invariant measure
$N\,du\,dv\,dP_u\,dP_v$ with the density
$$
\begin{aligned}
N=\frac{2uv+(u+v)(2d+\alpha_1)+\alpha_2-d\alpha_1}{\sqrt{\det({\bf J}-d{\boldsymbol n}\otimes{\boldsymbol n})}}
&(4\alpha_3+2\alpha_1\alpha_2-\alpha_1^3-d\alpha_1^2+\\
+&(\alpha_1^2-2\alpha_2+4d\alpha_1)(u+v)-4d(u+v)^2)^{-1},
\end{aligned}
$$
where $\alpha_1=\sum J_i,\,\alpha_2=\sum J_i^2,\alpha_3=J_1J_2J_3$.

According to the Chaplygin theory of reducing multiplier (\cite{9}), after the time reparameterization
$N(u,\,v)\,dt=d\tau$ the system ~\eqref{4} is transformed to the Lagrange form
\begin{equation}
\label{reduced}
\frac{d}{d\tau} \frac{\partial T}{\partial u'} -\frac{\partial T}{\partial u}=0,\quad
\frac{d}{d\tau}\frac{\partial T}{\partial v'} -\frac{\partial T}{\partial v}=0,
\qquad u'=\frac{du}{d\tau},\quad v'=\frac{dv}{d\tau}.
\end{equation}
As a result, under the time reparameterization we obtain an integrable Hamiltonian system on the cotangent
bundle $T^*S^2$ with local coordinates $u,v,p_u=\partial T/\partial u', p_v = \partial T/ \partial v'$.

\subsection*{A2. Separation of variables}
Equations \eqref{reduced} possess 2 homogeneous quadratic integrals, which come from $H,F_1$ \eqref{2} and which
can be written in the form
$$
\begin{aligned}
H=T & =\frac{1}{2}(g_{uu}(u')^2+2g_{uv}u'v'+g_{vv}(v')^2),\\
F_1 & =\frac{1}{2}(G_{uu}(u')^2+2G_{uv}u'v'+ G_{vv} (v')^2) .
\end{aligned}
$$
Explicit expressions for the coefficients $g_{uu}, \dots, G_{vv}$ as functions of $u,v$ are suppressed due to their
complexity.

According to the result of Eisenhart \cite{4} (see its modern accounting in \cite{Sokolov}),
separating variables $s_1, s_2$ can be chosen as the roots of the equation
\begin{equation}
\label{N1}
\det({\bf G}-s{\bf g})=0,
\end{equation}
$$
{\bf G}=\begin{Vmatrix}
      G_{uu} & G_{uv} \\
      G_{uv} & G_{vv} \\
    \end{Vmatrix},
{\bf g}=\begin{Vmatrix}
      g_{uu} & g_{uv} \\
      g_{uv} & g_{vv} \\
    \end{Vmatrix}.
$$

Note that the roots depend only on the local coordinates $u,v$ on the configuration space~$S^2$.

The spheroconical coordinates \eqref{333} depend explicitly on the roots $s_1, s_2$ as follows
$$
u=-\frac{1}{2}(y+\sqrt{y^2-4x}),\quad
v=-\frac{1}{2}(y-\sqrt{y^2-4x}),
$$
where
\begin{gather}
x=\frac{\pm s_2Q(s_1)\pm s_1Q(s_2)}{4d(s_1-s_2)},\quad
y=\frac{\pm Q(s_1)\pm Q(s_2)}{2d(s_1-s_2)},\notag\\
\label{N1_2}
Q(s)=\sqrt{b_1s^2-b_2s+b_3^2},\\
b_1=\frac{1}{16}(\Tr({\bf JA}))^2-\frac{1}{2}d\det {\bf A},\quad
b_3=2\det {\bf J}-\frac{1}{2}d\Tr({\bf JA}),   \label{b13} \\
b_2=\det {\bf J}\Tr({\bf JA})-\frac{1}{4}d(\Tr({\bf JA}))^2-\frac{1}{2}d\det{\bf A}\Tr{\bf J}
+d^2\det {\bf A}. \label{b2}
\end{gather}

In the new variables $s_1, s_2$ and the conjugated momenta $\displaystyle p_1=\frac{\partial T}{\partial s'_1},
p_2=\frac{\partial T}{\partial s'_2}$ the integrals take the Liouville form
\begin{equation}
\label{9}
H =\frac{S_1(s_1)}{s_1-s_2}p_1^2-\frac{S_2(s_2)}{s_1-s_2}p_2^2,\quad
F_1 =\frac{s_2S_1(s_1)}{s_1-s_2}p_1^2-\frac{s_1S_2(s_2)}{s_1-s_2}p_2^2,
\end{equation}
where
\begin{align}
\label{10}
S(x) & =\frac{2(8x^3+8(d-\epsilon)x^2+(2\epsilon^2\beta-4d\epsilon)x-4\gamma-d\beta+
\sqrt{\Lambda (x) })}{\gamma(2x-\epsilon+2d)^2}, \\
\Lambda & =x^2(\beta^2+8\alpha d)+2x(4\beta\gamma+d\beta^2-2d\alpha\epsilon+4\alpha d^2)+
(4\gamma+d\beta)^2 , \nonumber
\end{align}
and we used the notation
\begin{equation}
\label{notation}
\begin{aligned}
&\alpha=(J_2+J_1-J_3)(-J_2+J_2-J_3)(-J_2+J_2+J_3)=-A_1 A_2 A_3, \\
&\beta=J_1^2+J_2^2+J_3^2-2J_1J_2-2J_2J_3-2 J_3 J_1 = -A_1 A_2- A_2 A_3- A_3 A_1 , \\
&\gamma=J_1J_2J_3,\quad \epsilon=J_1+J_2+J_3.
\end{aligned}
\end{equation}

Due to the Hamilton equations with the Hamiltonian $H$ in \eqref{9}, the evolution of $s_1, s_2$ is descibed as follows
\begin{gather}
s_1' = \frac {\sqrt{2/\gamma} }{2 s_1- e+2d } \frac {y_1 }{ s_1-s_2}, \quad
s_2' = \frac {\sqrt{2/\gamma} }{2 s_2- e+2d } \frac {y_2 }{ s_2-s_1}, \label{quad3} \\
y_i = \sqrt {Y(x_i)} , \nonumber  \\
Y(x)=\Lambda (x)\cdot (h x-f) \left[ 8 x^3+8(d-e)x^2+(2 e^2-\beta-4 d e)x- c + \sqrt{\Lambda (x) }\right ] ,
\label{xy}
\end{gather}
where $h,f$ are the constants of the integrals $H, F_1$.

Hence, we performed a separation of variables, however the evolution equations \eqref{quad3} have a quite tedious form.

One can show that the equation $y^2 = Y(x)$ defines an algebraic curve of genus 2 on the plane
${\mathbb C}^2=(x,y)$.
According to the theory of algebraic curves, 
any curve of genus 2 is hyperelliptic and can be transformed to a canonical Weierstrass form by an appropriate
birational transformation of the coordinates $x,y$.

One of such transformations is induced by the chain of substitutions $x \to \xi \to z $
$$
x =\frac{4b_3\xi}{(\xi+b_2)^2-4b_1 b^2_3}, \quad
\xi= \frac {-(4 \det {\bf J}- d \Tr(\bf {A J} ) )\, z + \Tr\bf {J}-2d }{2\det {\bf A} \det {\bf I}},
$$
$b_1, b_2, b_3$ being defined in \eqref{b13}, \eqref{b2}. It converts $\Lambda(x)$ in \eqref{xy},
as well as $Q(x)$ \eqref{N1_2} into full squares.

After some tedious calculations, one finds that
in the new variables $z_1=z(x_1), z_2=z(x_2)$ the expressions \eqref{333} take the form \eqref{33}, which
ensures the reduction to hyperelliptic quadratures in the canonical form \eqref{N4}.


\subsection*{Acknowledgments} A.V.B. and I.S.M. research was partially supported by the Russian Foundation of Basic Research
(projects Nos. 08-01-00651 and 07-01-92210). I.S.M. also acknowledges the support from the RF
Presidential Program for Support of Young Scientists (MD-5239.2008.1). Yu.N.F. acknowledges the
support of grant BFM 2003-09504-C02-02 of Spanish Ministry of Science and Technology.

\end{document}